\documentstyle[12pt]{article}

\newcommand{\newsection}{    
\setcounter{equation}{0}
\section}
\def\appendix#1{
  \addtocounter{section}{1}
  \setcounter{equation}{0}
  \renewcommand{\thesection}{\Alph{section}}
  \section*{Appendix \thesection\protect\indent \parbox[t]{11.715cm} {#1} }
  \addcontentsline{toc}{section}{Appendix \thesection\ \ \ #1}
  }
\newcommand{\tr}[1]{\,{\rm tr}\,#1}
\newcommand{\Sp}[1]{\,{\rm Sp}\,#1}
\newcommand{\ntr}[1]{\,\frac {\rm tr}{N}\,#1}
\def\e{{\,\rm e}\,}

\def\eop{\vspace*{\fill}\pagebreak}
\def\be{\begin{equation}}
\def\ee{\end{equation}}
\def\bea{\begin{eqnarray}}
\def\eea{\end{eqnarray}}
\def\LA{\left\langle}
\def\RA{\right\rangle}

\newcommand{\Co}{\,\hbox{Cont}_\nu\,}
\newcommand{\rf}[1]{(\ref{#1})}
\newcommand{\eq}[1]{Eq.~(\ref{#1})}

\def\a{\alpha}

\def\d{\partial}

\def\l{\lambda}

\def\om{\omega}
\newcommand{\ie}{{\it i.e.}\ }

\newcommand{\p}{{\prime}}
\newcommand{\ra}{\rightarrow}
\newcommand{\eps}{\varepsilon}

\newcommand{\non}{\nonumber \\*}
\let\la=\lambda

\newcommand{\VVp}{{\cal V}^\prime}

\def\fun#1#2{\lower3.6pt\vbox{\baselineskip0pt\lineskip.9pt
\ialign{$\mathsurround=0pt#1\hfil##\hfil$\crcr#2\crcr\sim\crcr}}}

\begin{document}

\begin{titlepage}
\begin{flushright}
ITEP-YM-5-93 \\
SMI-93-5 \\
August, 1993
\end{flushright}
\vspace{.5cm}

\begin{center}
{\LARGE Adjoint Fermion Matrix Models}
\end{center} \vspace{1cm}
\begin{center} {\large Yu.\ Makeenko}
\footnote{E--mail:  \ makeenko@vxitep.itep.msk.su \ /
\ makeenko@nbivax.nbi.dk \ / \
makeenko@vxdesy.desy.de \ }
\\ \mbox{} \\
{\it Institute of Theoretical and Experimental Physics,}
\\ {\it B. Cheremushkinskaya 25, 117259 Moscow, RF} \\
\vspace{0.5cm} \mbox{} \\ {\large and} \\
\vspace{0.5cm} \mbox{} \\
{\large K.\ Zarembo} \\
 \mbox{} \\ {\it Steklov Mathematical Institute,} \\
{\it Vavilov st. 42, GSP-1, 117966 Moscow, RF}
\end{center}

\vskip 1 cm
\begin{abstract}
We study fermionic one-matrix, two-matrix and $D$-dimensional gauge invariant
matrix models.  In all cases we derive loop equations which unambiguously
determine the large-$N$ solution.  For the one-matrix case the solution is
obtained for an arbitrary interaction potential and turns out to be equivalent
to the one for the Hermitean one-matrix model with a logarithmic potential
and, therefore, belongs to the same universality class.  The explicit solutions
for the fermionic two-matrix and $D$-dimensional matrix models are obtained at
large $N$ (or in the spherical approximation) for the quadratic potential.

\end{abstract}

\vspace{1cm}
\noindent

\eop
\end{titlepage}
\setcounter{page}{2}

\section{Introduction}

Matrix models are usually associated with discretized random surfaces and
$2D$ quantum gravity~\cite{Kaz85}. The simplest Hermitean one-matrix model
corresponds to pure gravity~\cite{ds}
while a chain of Hermitean matrices
describes $2D$ gravity interacting with $c\leq1$ matter
\mbox{\cite{2,3,4}}.  A
natural multi-dimensional extension of this construction has been proposed
recently~\cite{KM92} in connection with induced lattice gauge theories.
However, the Hermitean matrix models possess, as is well-know, the $c=1$
barrier above which the stringy phase does not exist. For this reason it is
interesting to find out what happens for alternative matrix models.

In the present paper we consider {\it fermionic}\/ matrix models which involve
matrices with anticommuting elements. The simplest one --- the fermionic
one-matrix model --- is defined by the partition function
\be
Z=\int d\Psi d\bar{\Psi}\e^{-N\tr{}V(\sqrt{\bar{\Psi}\Psi})},
\label{1partition}
\ee
where $\Psi$ and $\bar{\Psi}$ are the $N\times N$ matrices whose
matrix elements are independent anticommuting Grassmann variables and $V$
stands
for a generic even potential
\be
V(\sqrt{\bar{\Psi}\Psi})=\sum_{k=0}^{\infty}g_{k}(\bar{\Psi}\Psi)^{k}\,.
\label{1potential}
\ee

A $D$-dimensional extension of the model~\rf{1partition} ---
the {\it adjoint fermion model} (AFM) --- is defined by
the partition function~\cite{KhM92b}
\be
Z_{AFM}=\int \prod_{x,\mu} dU_{\mu}(x) \prod_x  d\Psi_x
d\bar{\Psi}_x \e^{-S_{F}[\Psi,\bar{\Psi},U]}
\label{fpartition}
\ee
with $x$ labeling the sites of a $D$-dimensional lattice.
Here $S_{F}[\Psi,\bar{\Psi},U]$ is the lattice fermion action
\bea
 & & S_{F}[\Psi,\bar{\Psi},U] =
 \sum_x N \tr{\Big(V(\sqrt{\bar{\Psi}_x \Psi_x})
\nonumber \\* & & - c
\sum_{\mu=1}^D [\bar{\Psi}_x P_\mu^-U_\mu(x)\Psi_{x+\mu}U_\mu^\dagger(x)}
+\bar{\Psi}_{x+\mu} P_\mu^+U_\mu^\dagger(x)\Psi_x U_\mu(x)]\Big)
\label{faction}
\eea
with
\be
P_\mu^\pm=r\pm\gamma_\mu \label{projectors}
\ee
being the projectors. The case $r=0$ corresponds to chiral fermions
while $r=1$ is associated with Wilson fermions. As is well-known,
the chiral fermions describe $2^D $ flavors in the naive continuum
limit while Wilson fermions are associated with $1$ flavor.

For $D\leq1$, which is associated with the case of a matrix chain, the gauge
field $U_\mu(x)$ can be absorbed by a (local) gauge transformation of $\Psi_x$
and $\bar{\Psi}_x$
so that the model~\rf{fpartition} reduces to fermionic multi-matrix models.
The two-matrix model is associated, in particular, with $D=1/2$.
In contrast to their bosonic counterparts, the fermionic matrix models are not
studied and the proper critical index $\gamma_{string}$ is not calculated.
This is one of the motivations for studies of the fermionic matrix models.

Another motivation for studies of AFM is a recent interest in the problem
of inducing QCD at large $N$ which is caused by Ref.~\cite{KM92}.
There are some important features which differ AFM
from its Hermitean analogue
which is known as the Kazakov--Migdal model:
\begin{itemize} \addtolength{\itemsep}{-7pt} \vspace{-9pt}
\item[i)] There is no asymptotic freedom of the gauge coupling constant for
AFM with chiral fermions or Kogut--Susskind fermions, the latter are
associated with
$4$ flavors in the continuum.
For this reason
the kinetic term of the gauge field is not essential similarly to the case of
quantum electrodynamics~\cite{Zel67}.
\item[ii)] AFM with the quadratic potential has no instability at any
values of $m$.
\item[iii)] AFM with the quadratic potential undergoes~\cite{KhM92b} a
first order large-$N$ phase transition with decreasing $m$ which is associated
with the restoration of area law.
\item[iiii)]
The matrices $\Psi$ can not be diagonalized by a gauge transformation so that
saddle point methods can not be applied straightforwardly at large $N$.
We shall find, however, a large-$N$ solution of AFM by means of loop
equations.
\vspace{-7pt} \end{itemize}

In the present paper we derive the set of loop equations which determines
the large-$N$ solution of AFM with an arbitrary potential at any $D$.
For the fermionic one-matrix model we find the solution explicitly and show
that it coincides to any order of the $1/N$-expansion with that of the
Hermitean one-matrix model with a logarithmic potential.
These two models belong, therefore, to the same universality class.
For the fermionic
two-matrix model we find explicitly the large-$N$ solution for the quadratic
potential~\rf{1potential} and obtain an algebraic equation of degree $2J$ which
determines the solution for the polynomial potential of degree $2J$.
We consider the loop equations of the two-matrix model to any order of
the $1/N$-expansion and discuss their relation to integrable hierarchies.
For the $D$-dimensional AFM we obtain the set of equations which
determines the large-$N$ solution for an arbitrary potential and write down the
explicit solution in the case of the quadratic potential.

\newsection{Fermion one-matrix model}

\subsection{Loop equation for fermion one-matrix model \label{1mmle}}

The simplest example of fermionic matrix models is the one-matrix model
defined by the partition function~\rf{1partition}.
The connected correlators $\left\langle\frac{1}{N}\tr{(\bar{\Psi}
\Psi)^{k_{1}}}\ldots\frac{1}{N}\tr{(\bar{\Psi}
\Psi)^{k_{m}}}\right\rangle_{conn.}$,
where the average is defined with the same measure as in~\rf{1partition},
 can be obtained differentiating
$\log Z$ w.r.t. $g_{k_{1}},\ldots,g_{k_{m}}$. Note that  the
 correlators with sufficiently large $k_{1},\ldots,k_{m}$ vanish for finite $N$
identically due to nilpotence of Grassmann variables. We introduce the
generating functions (loop correlators)
\be
W(\l_{1},\ldots,\l_{m})=\left\langle\ntr{\frac{\l_{1}}
{\l_{1}^{2}-\bar{\Psi}\Psi}}\ldots\ntr{\frac{\l_{m}}
{\l_{m}^{2}-\bar{\Psi}\Psi}}\right\rangle_{conn.}
\label{multiloop}
\ee
which are understood as the power series in
$\l_{1}^{-2},\ldots,\l_{m}^{-2}$ (for finite $N$ they are polynomials).
The multi-loop correlators can be obtained from the partition function by
application of the loop insertion operator
\be
W(\l_{1},\ldots,\l_{m})=\frac{1}{N^{2}}\frac{\delta}{\delta V(\l_{1})}
\ldots\frac{1}{N^{2}}\frac{\delta}{\delta V(\l_{m})}\log Z,
\label{1loopinsertion}
\ee
where
\be
\frac{\delta}{\delta V(\l)}=-\sum_{k=0}^{\infty}\l^{-2k-1}\frac{\partial}
{\partial g_{k}}.
\label{insertion}
\ee

The loop equation can be derived using the invariance of the integral
\be
\int d\Psi d\bar{\Psi}\e^{-N\tr{}V(\sqrt{\bar{\Psi}\Psi})}
\; \Psi\frac{\l} {\l^{2}-\bar{\Psi}\Psi} = 0
\ee
under an infinitesimal shift $\Psi\ra\Psi+\xi$ and reads
\be
\int_{C_{1}}\frac{d\om}{4\pi i}\frac{V'(\om)}{\l-\om}W(\om)=W^{2}(\l)-
\frac{2}{\l}W(\l)+\frac{1}{N^{2}}\frac{\delta}{\delta V(\l)}W(\l).
\label{1loopeq}
\ee
The contour $C_{1}$ encircles anticlockwise the singularities of $W(\om)$.
Eq.(\ref{1loopeq}) is supplemented with the asymptotic condition
\be
W(\l)\ra \frac 1\l ~~~~ {\rm as}~~ \l\ra\infty,
\label{1as}
\ee
which is a consequence of the definition (\ref{multiloop}). Notice that one
obtains the single functional equation for $W(\l)$. This is due to the
fact that $\tr{}V(\sqrt{\bar{\Psi}\Psi})$ contains a complete set of
operators. The $m$-loop correlator can be obtained from $W(\l)$ by
$(m-1)$-fold application of $\delta/\delta V(\l_{i})$ according to
(\ref{1loopinsertion}).

The loop equation (\ref{1loopeq}) can be represented as a set of Virasoro
constraints imposed on the partition function. The coefficients of
expansion of \eq{1loopeq} in $1/\l^{2}$ can be rewritten, using
(\ref{1loopinsertion}) and (\ref{insertion}), as
\be
L_{n}Z=0,~~~~~n\geq 0
\label{1Virasoroconstraints}
\ee
where the operators
\be
L_{n}=\sum_{k=0}^{\infty}kg_{k}\frac{\d}{\d g_{n+k}}+\frac{1}{N^{2}}
\sum_{k=0}^{n}\frac{\d^{2}}{\d g_{k}\d g_{n-k}}+2\frac{\d}{\d g_{n}}
\label{1Vir}
\ee
satisfy Virasoro algebra
\be
[L_{n},L_{m}]=(n-m)L_{n+m}.
\label{Virasoro}
\ee
The Virasoro constraints~\rf{1Virasoroconstraints} look similar to the ones
for the complex one-matrix model (see Ref.~\cite{Mak91} for a review) where
the last term on the r.h.s.\ of \eq{1Vir} was absent.

\subsection{Solution at $N=\infty$ \label{1sol}}

The loop equation (\ref{1loopeq}) can be solved order by order in $1/N^2$.
 The third term on the r.h.s.\ is suppressed by the factor $1/N^{2}$.
 Therefore, one can omit it as $N\ra\infty$ so that one gets
\be
\int_{C_{1}}\frac{d\om}{4\pi i}\frac{V'(\om)}{\l-\om}W(\om)=W^{2}(\l)-
\frac{2}{\l}W(\l)\,.
\label{1le}
\ee
 Doing the integral on the
 l.h.s.\ by calculation of the residues at $\om=\l$ and at $\om=\infty$, one
 obtains the quadratic equation for $W(\l)$ whose solution reads
 \be
 W(\l)=\frac{1}{4}V'(\l)+\frac{1}{\l}-\frac{1}{\l}\sqrt{\left(\frac{1}{4}\l
 V'(\l)+1\right)^{2}-\l^{2}Q(\l)},
 \label{multicutsolution}
 \ee
 where $Q(\l)$ is an even polynomial of degree $2J-2$ if
  $V(\l)$ is the one of degree $2J$.
 The coefficients of $Q(\l)$ are unambiguously determined by imposing the
analytic structure of $W(\l)$ and by the asymptotic condition (\ref{1as}).

The explicit solution for the simplest Gaussian potential is
\be
 W(\l)=\frac{1}{2}g_{1}\l+\frac{1}{\l}-\frac{1}{\l}\sqrt{\frac{1}{4}
 g_{1}^{2}\l^{4} +1} \,.
 \label{1Gauss}
 \ee
 This solution as a function of $\l^2$ has one cut along imaginary axis
 which represents the `support of the
 eigenvalue density' of the matrix ${\bar{\Psi}\Psi}$.  Note that this
cut does not lie on the real axis. As it is shown below, this is the general
 property of the one-cut solution. Another peculiar property of the
fermionic Gaussian model is that $\langle\ntr{(\bar{\Psi}\Psi)}^{2k}\rangle=0$
to the leading order of the $1/N$-expansion.
The fact that  $\langle\ntr{}\bar{\Psi}\Psi\bar{\Psi}\Psi\rangle$ vanishes
can be easily verified using, for example,
the Wick theorem which results in two terms with opposite signs.

 The general solution (\ref{multicutsolution}) has $n$ cuts
 on the complex $\l^2$ plane
 with $n\leq J$ for $V(\l)$ being the polynomial of degree $2J$.
The one-cut solution
can be represented in the following form
\be
W(\l)=\int_{C_{1}}\frac{d\om}{4\pi
    i}\frac{\frac{1}{2}V'(\om)+\frac{2}{\om}}{\l-\om}
    \sqrt{\frac{(\l^{2}-x)(\l^{2}-y)}{(\om^{2}-x)(\om^{2}-y)}},
\label{2cut}
\ee
The ends of the cut are determined by the asymptotic condition (\ref{1as})
which yields
\bea
\int_{C_{1}}\frac{d\om}{4\pi
    i}\frac{\frac{1}{2}V'(\om)+\frac{2}{\om}}{\sqrt{(\om^{2}-x)
    (\om^{2}-y)}}=0\,,
\nonumber\\
\int_{C_{1}}\frac{d\om}{4\pi
    i}\frac{\frac{1}{2}\om^{2}V'(\om)+2\om}{\sqrt{(\om^{2}-x)
    (\om^{2}-y)}}=1\,.
\label{ends}
\eea

The solution~\rf{2cut}, \rf{ends} reminds the simplest one-cut solution
(as a function of $\l^2$) of the Hermitean one-matrix model.
This is not a coincidence.
As is shown in Appendix~A, the fermionic one-matrix model~\rf{fpartition}
is {\it equivalent}\/
to all orders of the $1/N$-expansion
to the Hermitean one-matrix model
with the logarithmic potential
\be
{V}^{\bf H}(\Phi) = V(\sqrt{\Phi}) +2\log{\Phi}
\ee
where $\Phi=\bar{\Psi}\Psi$ is the $N\times N$ Hermitean matrix.
The large-$N$ solution which is associated with the fermionic model is,
however, somewhat different as compared with the standard solutions of the
Hermitean one-matrix models, because it is known that the integer
value of coupling constant in front of the logarithm is critical~\cite{Tan91}
for the large-$N$ limit. In particular,
the points $x$ and $y$ cannot lie on the real axis. Otherwise the
solution were have a pole at zero, so that the contour $C_{1}$ would
encircle the origin.
One can readily verify that Eqs.~(\ref{ends}) would have no solution in this
case.

The multi-loop correlators can be calculated using the equation
 (\ref{1loopinsertion}). For the two-loop correlator we have
 \be
 W(\nu,\l)=\frac{1}{N^{2}}\frac{\l \nu}{4(\l^{2}-\nu^{2})^{2}}\left[\frac
 {(\l^{2}-x)(\nu^{2}-y)+(\l^{2}-y)(\nu^{2}-x)}
 {\sqrt{(\l^{2}-x)(\l^{2}-y)(\nu^{2}-x)(\nu^{2}-y)}}
 -2\right]
 \label{1twoloop}
 \ee
which looks like the two-loop correlator~\cite{AJM90} of the Hermitean
one-matrix model.

Eqs.~(\ref{ends}) simplify for reduced potentials with $g_{2k}=0$ which
change sign under the chiral transformation $\bar\Psi\Psi\ra-\bar\Psi\Psi$.
Then for $y=-x$ the first equation  reads
\be
w(x)=2,~~w(x)\equiv\sum_{k} (-)^kg_{2k+1}\frac{(2k+1)!}{2^{2k}(k!)^{2}}
x^{2k+1}
\label{1w}
\ee
while the second one turns into identity.  \eq{1w}
coincides with the proper one for the Hermitean one-matrix model with the
polynomial potential
\be
U(\Phi) = \sum_k (-)^k g_{2k+1} \Phi^{2k} \,.
\ee
The critical behavior emerges at the point where
\be
w'(x_{c})=\ldots
=w^{(m-1)}(x_{c})=0\,,~~~~w(x_{c})=2~.
\ee
This is nothing but the standard
multi-critical point of the Hermitean one-matrix model.
For this reason the critical index $\gamma_{string}$
coincides in genus zero with the one of
$2D$ gravity:
\be
\gamma_{string} = - \frac 12 \,.
\ee

\newsection{Fermion two-matrix model \label{2le}}

\subsection{Derivation of loop equations}

Let us define the partition function of the fermion two-matrix model by
\be
Z_{2}=\int d\Psi_1 d\bar{\Psi}_1 d\Psi_2 d\bar{\Psi}_2
\e^{N\tr{}\left(-{\cal V}(\sqrt{\bar{\Psi}_1\Psi_1})
-\tilde{\cal V}(\sqrt{\bar{\Psi}_2\Psi_2})+
c [- \bar{\Psi}_1\Psi_2 + \bar{\Psi}_2\Psi_1 ]\right) }\,.
\label{2partition}
\ee
The asymmetric case is associated with non-equal
potentials
\be
{\cal V}(\sqrt{\bar{\Psi}\Psi}) = \sum_k t_k (\bar{\Psi}\Psi)^k
\label{2potential}
\ee
and
\be
\tilde{\cal V}(\sqrt{\bar{\Psi}\Psi}) =
\sum_k \tilde t_k (\bar{\Psi}\Psi)^k  \,.
\label{2potentialt}
\ee

Let us define the odd-odd and even-even two-point correlators,
respectively, by
\be
{G}(\nu,\l) =
\left\langle \ntr{}
\Big(  \Psi_1 \frac{1}{(\nu^2-\bar{\Psi}_{1}\Psi_{1})}
\frac{1}{(\l^2-\bar{\Psi}_{2}\Psi_{2})} \bar{\Psi}_{2} \Big)
 \right\rangle\,,
\label{2defG}
\ee
and
\be
W(\nu,\l)=
\left\langle \ntr{}
\Big( \frac{\nu}{(\nu^2-\bar{\Psi}_{1}\Psi_{1})}
\frac{\l}{(\l^2-\bar{\Psi}_{2}\Psi_{2})}  \Big)
 \right\rangle\,.
\label{2defH}
\ee
Analogously, the correlators of arbitrary powers of $\bar{\Psi}_{1}\Psi_{1}$
are determined by
\be
W_0(\nu)=
\left\langle \ntr{}
\Big( \frac{\nu}{\nu^2-\bar{\Psi}_{1}\Psi_{1}} \Big)
 \right\rangle
\label{2defE}
\ee
which enters the asymptotic expansion of $W(\nu,\l)$ in
$1/\l$:
\be
W(\nu,\l)= \sum_{n=0}^\infty \frac{W_{2n}(\nu)}{\l^{2n+1}}  ~.
\label{2bcH}
\ee
Similarly, we define
\be
G(\nu,\l)= \sum_{n=0}^\infty \frac{G_{2n+1}(\nu)}{\l^{2n+2}}  ~.
\label{2bcG}
\ee

The first loop equation results from the invariance of the measure in
\be
\left\langle \ntr{}
\Big( t^A  \frac{1}{(\nu^2-\bar{\Psi}_{1}\Psi_{1})}
\frac{1}{(\l^2-\bar{\Psi}_{2}\Psi_{2})} \bar{\Psi}_{2}
 \Big)  \right\rangle = 0
\ee
under the shift
\be
\left(\bar{\Psi}_{1}\right)_{kl} \ra \left(\bar{\Psi}_{1}\right)_{kl} +
\eps^{A}_1 \left(t^A\right)_{kl}
\ee
where $t^A$ ($A=1,\ldots,N^2$) are the generators of the $U(N)$ which are
normalized by
\be
\sum_{A=1}^{\,N^2} [t^A]_{ij} [t^A]_{kl} = N \delta_{il} \delta_{kj}~.
\label{completeness}
\ee
Quite similarly to Ref.~\cite{2mamo} which deals with the Hermitean
two-matrix model,  this equation takes
at $N=\infty$, when factorization holds, the form
\be
 \int_{C_1} \frac{d \om}{4\pi i}
\frac{\VVp(\om)}{\nu - \om}G(\om, \la)=
W_0(\nu) G(\nu, \la) + c  \Big[\la W(\nu, \la) - W_0(\nu) \Big]
\label{2main1}
\ee
where the contour $C_1$ encircles anticlockwise the singularities of
$G(\om,\la)$ as a  function of $\om$.

Since \eq{2main1} expresses $G$ via $W$, one needs one more equation which
relates these quantities. It can be derived making the shift
\be
\left({\Psi}_{1}\right)_{kl} \ra \left({\Psi}_{1}\right)_{kl} +
\eps^{A}_1 \left(t^A\right)_{kl}
\ee
in
\be
\left\langle \ntr{}
\Big( t^A \Psi_1  \frac{\nu}{(\nu^2-\bar{\Psi}_{1}\Psi_{1})}
\frac{\l}{(\l^2-\bar{\Psi}_{2}\Psi_{2})}  \Big) \right\rangle = 0 \,.
\ee
The resulting equation reads
\be
 \int_{C_1} \frac{d \om}{4\pi i}
\frac{\VVp(\om)}{\nu - \om}W(\om, \la)=W_0(\nu) W(\nu, \la)
- 2 \frac{1}{\nu} W (\nu, \l)
 - c \l  G(\nu, \la)  \,.
\label{2main2}
\ee

\subsection{Exact solution at $N=\infty$}\label{2exact}

Quite similarly to the scalar case~\cite{DMS93},
Eqs.~\rf{2main1}, \rf{2main2} can be rewritten as an equation for $W_0(\nu)$.
To this aim let us express $W_{2n}(\nu)$ and $G_{2n+1}(\nu)$, which are defined
by Eqs.~\rf{2bcH} and \rf{2bcG}, for $n\geq1$ via $W_0(\nu)$:
\bea
- c G_{2n+1}(\nu) = \int_{C_1} \frac{d \om}{4\pi i}
\frac{\VVp(\om)}{\nu - \om}W_{2n}(\om) +\left(\frac 2\nu - W_0(\nu) \right)
W_{2n}(\nu)~, \non
c W_{2n+2}(\nu) = \int_{C_1} \frac{d \om}{4\pi i}
\frac{\VVp(\om)}{\nu - \om}G_{2n+1}(\om) - W_0(\nu) G_{2n+1}(\nu)~.
\label{recurrence}
\eea
These recurrence relations are obtained by expanding
Eqs.~\rf{2main1}, \rf{2main2} in $1/\l$.

The equation for $W_0$ can now be obtained taking the $1/\nu$ term of the
expansion of \eq{2main1} (with 1 and 2 interchanged) in $1/\nu$ and reads
\be
\sum_k k \tilde t_k G_{2k-1}(\l) = c [\l W_0(\l) -1 ]~.
\label{2smart}
\ee
The equations which come from the next terms of the expansion of
Eqs.~\rf{2main1} and  \rf{2main2} in $1/\nu$ should be
automatically satisfied as a
consequence of \eq{2smart}.
For the potential $\tilde{\cal V}$ being the polynomial of the highest power
$2J$, \eq{2smart} is a $2J$-ic algebraic equation for $W_0(\l)$
similarly to the scalar case~\cite{2mamo,DMS93}.

It is instructive to consider in detail the case of the quadratic symmetric
potentials
\be
{\cal V}(\om) =\tilde {\cal V}(\om) = t_1 \om^2~~~~~~~~~\hbox{(quadratic
potential)}
\ee
when
\be
- c G_1(\l) = t_1 (\l W_0(\l)-1) +\left(\frac 2\l - W_0(\l) \right)
W_{0}(\l)
\label{2quadr}
\ee
and \eq{2smart} is quadratic:
\be
t_1 G_{1}(\l) = c [\l W_0(\l) -1 ]~.
\label{2quasmart}
\ee
The solution to (\ref{2quadr}), (\ref{2quasmart}) has the same form as the
 one (\ref{1Gauss}) for the one-matrix model with
\be
g_{1}=t_{1}+\frac{c^{2}}{t_{1}} \,.
\label{2g1}
\ee

\subsection{Relation to integrable hierarchies}

While Eqs.~\rf{2main1}, \rf{2main2} have been derived at $N=\infty$, they can
be extended for the two-matrix model to any order of the $1/N$-expansion. The
difference between $N=\infty$ and finite-$N$ equations resides in irreducible
correlators (of order $1/N^2$) which can be conveniently expressed in the
asymmetric case ${\cal V}(\nu) \neq \tilde{\cal V}(\nu)$ via the loop insertion
operator
\be
\frac{\delta}{\delta {\cal V}(\nu)} \equiv - \sum_{k=0}^\infty
\frac {\partial}{\partial t_k} \frac {1}{\nu^{k+1}} \,.
\label{loopinsertion}
\ee

The set of the loop equations which extends Eqs.~\rf{2main1}, \rf{2main2}
 to finite $N$ reads
\be
 \int_{C_1} \frac{d \om}{4\pi i}
\frac{\VVp(\om)}{\nu - \om}G(\om, \la)=
W_0(\nu) G(\nu, \la)
+\frac {1}{N^2} \frac{\delta}{\delta {\cal V}(\nu)}  G(\nu, \la)
+ c  \Big[\la W(\nu, \la) - W_0(\nu) \Big]
\label{2main1f}
\ee
and
\be
 \int_{C_1} \frac{d \om}{4\pi i}
\frac{\VVp(\om)}{\nu - \om}W(\om, \la)=W_0(\nu) W(\nu, \la)
+\frac {1}{N^2} \frac{\delta}{\delta {\cal V}(\nu)}  W(\nu, \la)
- 2 \frac{1}{\nu} W (\nu, \l)
 - c \l  G(\nu, \la)
\label{2main2f}
\ee
so that for the recurrence relations one gets
\bea
- c G_{2n+1}(\nu) = \int_{C_1} \frac{d \om}{4\pi i}
\frac{\VVp(\om)}{\nu - \om}W_{2n}(\om) +\left(\frac 2\nu - W_0(\nu) \right)
W_{2n}(\nu)-\frac {1}{N^2} \frac{\delta}{\delta {\cal V}(\nu)}
W_{2n}(\nu)\,, \non
c W_{2n+2}(\nu) = \int_{C_1} \frac{d \om}{4\pi i}
\frac{\VVp(\om)}{\nu - \om}G_{2n+1}(\om) - W_0(\nu) G_{2n+1}(\nu)
-\frac {1}{N^2} \frac{\delta}{\delta {\cal V}(\nu)} G_{2n+1}(\nu)
\,.
\label{recurrencef}
\eea

Introducing the analogues of the loop insertion
operator~\rf{loopinsertion} for $G_{2n+1}(\nu)$ and $W_{2n}(\nu)$
with $n\geq0$:
\bea
G_{2n+1}(\nu)=\frac {1}{Z_2} {\cal L}_{2n+1}(\nu) Z_2\,, \non
{\cal L}_{2n+1}(\nu) \equiv - \sum_{k=-n}^\infty
{\cal W}_k^{(2n+2)}\frac {1}{\nu^{2(k+n+1)}}
\label{delL_2n+1}
\eea
and
\bea
W_{2n}(\nu)=\frac {1}{Z_2} {\cal L}_{2n}(\nu) Z_2\,, \non
{\cal L}_{2n}(\nu) \equiv - \sum_{k=-n}^\infty
{\cal W}_k^{(2n+1)}\frac {1}{\nu^{2(k+n)+1}}
\label{delL_2n}
\eea
so that
\be
{\cal L}_0(\nu)=\frac{\delta}{\delta {\cal V}(\nu)}\,,
{}~~~~~~~{\cal W}_k^{(1)}=
\frac {\partial}{\partial t_k} \,,
\ee
it can be shown after a little algebra that
\eq{recurrence} is now identically satisfied for any $Z_2$ provided that
the operators ${\cal W}_k^{(n)}$ obey the recurrence relations
\bea
-c {\cal W}_k^{(2n+2)} = \sum_{m=1}^\infty m t_m {\cal W}_{m+k}^{(2n+1)} +
\frac {1}{N^2} \sum_{m=0}^{k+n}
\frac {\partial}{\partial t_m}{\cal W}_{k-m}^{(2n+1)}
+2 {\cal W}_{k}^{(2n+1)}
{}~~~~~~~k\geq -n  \non
c {\cal W}_k^{(2n+1)} = \sum_{m=1}^\infty m t_m {\cal W}_{m+k}^{(2n)} +
\frac {1}{N^2} \sum_{m=0}^{k+n-1}
\frac {\partial}{\partial t_m}{\cal W}_{k-m}^{(2n)}\,.
{}~~~~~k\geq -n
\label{recurrenceW}
\eea
Analogous operators for the Hermitean two-matrix model
were advocated by Marshakov {\it et al.}~\cite{2mamo}.

The set of equations similar to
Eqs.~\rf{2main1f} to \rf{recurrenceW} can be obtained by varying the partition
function~\rf{2partition} w.r.t.\ $\Psi_2$ rather than $\Psi_1$. One gets
\bea
\tilde{G}_{2n+1}(\l)= -\frac {1}{Z_2} \sum_{k=-n}^\infty
\tilde{\cal W}_k^{(2n+2)}\frac {1}{\l^{2(k+n+1)}} Z_2~, \non
\tilde{W}_{2n}(\l)= -\frac {1}{Z_2} \sum_{k=-n}^\infty
\tilde{\cal W}_k^{(2n+1)}\frac {1}{\l^{2(k+n)+1}} Z_2~,
\label{delbL_n}
\eea
with $\tilde{\cal W}^{(n)}_k$ given by the same formulas as
${\cal W}^{(n)}_k$ with $t_m$ replaced by $\tilde{t}_m$.

The equation for $\tilde{W}_0(\l)$ can be obtained taking the $1/\nu$ term of
the expansion of \eq{2main1f} in $1/\nu$ and reads
\be
\sum_{m\geq1} m {t}_{m}\tilde{G}_{2m-1}(\l) = c(\l \tilde{W}_0(\l) -1)~.
\label{bsmart}
\ee
Similarly, one gets
\be
\sum_{m\geq1} m \tilde{t}_{m}G_{2m-1}(\l) = c(\l W_0(\l) -1)
\label{2smartf}
\ee
which determines $W_0(\l)$ versus $\{t_k\}$ and  $\{\tilde{t}_k\}$.
\eq{2smartf} coincides with \eq{2smart} above.

If all $\tilde{t}_m$'s vanish except for some $m=n$ ($\tilde{t}_n=1/n$),
\eq{2smartf} reduces to the constraints
\be
{\cal W}^{(2n)}_k Z_2 = c {\cal W}^{(1)}_{k+n} Z_2 ~~~~~~~~~~~~~~~
k\geq 1-n
\ee
imposed on the partition function $Z_2$.

\newsection{$D$-dimensional AFM at large $N$ \label{Ddim}}

\subsection{Arbitrary potential}

Let us consider AFM on a $D$-dimensional lattice with is defined
by the partition function~\rf{fpartition}.
Let us define again the odd-odd and even-even one-link correlators,
respectively, by
\be
{G}^{ij}_{\mu}(\nu,\l) =
\left\langle \ntr{}
\Big(  \Psi^i_x \frac{1}{\nu^2-\bar{\Psi}_{x}\Psi_{x}}U_\mu(x)
\frac{1}{\l^2-\bar{\Psi}_{x+\mu}\Psi_{x+\mu}} \bar{\Psi}^j_{x+\mu}
U^\dagger_\mu(x) \Big)
 \right\rangle\,,
\label{defG}
\ee
where $i,j=1,\ldots, s$ are spinor indices, and
\be
W(\nu,\l)=
\left\langle \ntr{}
\Big( \frac{\nu}{\nu^2-\bar{\Psi}_{x}\Psi_{x}}U_\mu(x)
\frac{\l}{\l^2-\bar{\Psi}_{x+\mu}\Psi_{x+\mu}} U^\dagger_\mu(x) \Big)
 \right\rangle\,.
\label{defH}
\ee
Analogously, the correlators of arbitrary powers of $\bar{\Psi}_{x}\Psi_{x}$
at the same site $x$ are determined by
\be
W_0(\nu)=
\left\langle \ntr{}
\Big( \frac{\nu}{\nu^2-\bar{\Psi}_{x}\Psi_{x}} \Big)
 \right\rangle\,.
\label{defE}
\ee
The asymptotic expansion of $W(\nu,\l)$ in
$1/\l$ is defined again by \eq{2bcH}.

The first loop equation results from the invariance of the measure in
\be
\left\langle \ntr{}
\Big( t^A  \frac{1}{\nu^2-\bar{\Psi}_{x}\Psi_{x}}U_\mu(x)
\frac{1}{\l^2-\bar{\Psi}_{x+\mu}\Psi_{x+\mu}} \bar{\Psi}^j_{x+\mu}
U^\dagger_\mu(x) \Big)
 \right\rangle = 0
\ee
under the shift
\be
\left(\bar{\Psi}^i_{x}\right)_{kl} \ra \left(\bar{\Psi}^i_{x}\right)_{kl} +
\eps^{A\,i}_x \left(t^A\right)_{kl} \,.
\ee
Quite similarly to Ref.~\cite{DMS93} which deals with the Kazakov--Migdal
model,  this equation takes
at $N=\infty$, when factorization holds, the form of loop
equation~\rf{2main1} for the two-matrix AFM:
\be
 \int_{C_1} \frac{d \om}{4\pi i}
\frac{\VVp(\om)}{\nu - \om}G^{ij}_\mu(\om, \la)=
W_{0}(\nu) G^{ij}_\mu(\nu, \la) - c \frac 1s \left(P_\mu^\mp\right)^{ij}
\Big[\la W(\nu, \la) - W_0(\nu) \Big]
\label{main1}
\ee
with the potential ${\cal V}$  given by
\be
\VVp(\om)\equiv V^\prime(\om)-2c\sigma (2D-1) F(\om)
\label{defL}
\ee
where
\be
\sigma = r^2 -1\,.
\ee

The function
\be
F(\om)=\sum_{n=1}^\infty F_n \om^{2n-1}
\ee
in \eq{defL}
is  the one which appears in the one-link correlator
 \be
 \left\langle \ntr{} \Big(t^a U \chi U^\dagger\Big)\right\rangle_{o.l.}
\equiv \frac{\int d\chi
d\bar{\chi} dU\, \e^{-N\tr{}\left(V(\sqrt{\bar{\chi}\chi})- c[\bar{\psi}
P_\mu^-U \chi U^\dagger + \bar{\chi} P_\mu^+ U^\dagger \psi U]\right)}
\ntr{} \Big(t^aU \chi U^\dagger\Big)} {\int d\chi d\bar{\chi}
dU\,\e^{-N\tr{}\left(V(\sqrt{\bar{\chi}\chi})- c[\bar{\psi} P_\mu^-U \chi
U^\dagger + \bar{\chi} P_\mu^+ U^\dagger \psi U]\right)}}.
  \label{1link}
 \ee
where the averaging is only w.r.t.\ $U$ and $\chi$ while $\psi$ plays
the role of an external field.  Similarly to the case of the Hermitean
model~\cite{Mig92a,Mig92d}, the following formula holds at $N=\infty$:
\be
\left\langle \ntr{}\Big( t^A U \chi^i
U^\dagger\Big)\right\rangle_{o.l.}
= \left(P_\mu^+\right)^{ij} \sum_{n=1}^\infty  F_{n}
\ntr{}\left(t^A \psi^{j} (\bar{\psi}\psi)^{n-1} \right) \,,
\label{Lambda}
\ee
This formula is discussed in Appendix~B.

One more equation can be derived making the shift
\be
\left({\Psi}^j_{x}\right)_{kl} \ra \left({\Psi}^j_{x}\right)_{kl} +
\eps^{A\,j}_x \left(t^A\right)_{kl}
\ee
in
\be
\left\langle \ntr{}
\Big( t^A \Psi^i_x  \frac{\nu}{\nu^2-\bar{\Psi}_{x}\Psi_{x}}U_\mu(x)
\frac{\l}{\l^2-\bar{\Psi}_{x+\mu}\Psi_{x+\mu}}
U^\dagger_\mu(x) \Big) \right\rangle = 0 \,.
\ee
It takes the form of \eq{2main2}:
\be
 \int_{C_1} \frac{d \om}{4\pi i}
\frac{\VVp(\om)}{\nu - \om}W(\om, \la)=W_0(\nu) W(\nu, \la)
- (s+1) \frac{1}{\nu} W (\nu, \l)
 - c \l \left(P_\mu^\pm\right)^{ji}
 G^{ij}_\mu(\nu, \la)  \,.
\label{main2}
\ee

To solve the system~\rf{main1} and \rf{main2}, it is convenient to introduce
\be
\hat G (\nu,\l) \equiv \left( P^\pm_\mu\right)^{ji}
 G^{ij}_\mu(\nu, \la)
 \label{hatdefG}
\ee
and
\be
\hat G(\nu,\l)= \sum_{n=0}^\infty \frac{\hat G_{2n+1}(\nu)}{\l^{2n+2}}  ~.
\label{hatbcG}
\ee

Multiplying \eq{main1} by $( P^\pm_\mu )^{ji}$ one rewrites
Eqs.~\rf{main1} and \rf{main2} as
\be
\int_{C_1} \frac{d \om}{4\pi i}
\frac{\VVp(\om)}{\nu - \om}\hat G(\om, \la)=
W_0(\nu) \hat G(\nu, \la) - c\sigma \Big[\la W(\nu, \la) - W_0(\nu) \Big]
\label{hatmain1}
\ee
and
\be
 \int_{C_1} \frac{d \om}{4\pi i}
\frac{\VVp(\om)}{\nu - \om}W(\om, \la)=W_0(\nu) W(\nu, \la)
- (s+1) \frac{1}{\nu} W (\nu, \l)
 - c \l \hat G(\nu, \la)  \,.
\label{hatmain2}
\ee
At $s=1$ and $\sigma=-1$ these equations coincide with~\rf{2main1} and
\rf{2main2}.

The analogue of \eq{recurrence} now reads
\bea
- c \hat G_{2n+1}(\nu) = \int_{C_1} \frac{d \om}{4\pi i}
\frac{\VVp(\om)}{\nu - \om}W_{2n}(\om) +\left(\frac {s+1}{\nu} - W_0(\nu)
\right) W_{2n}(\nu)~, \non
-c\sigma W_{2n+2}(\nu) = \int_{C_1} \frac{d \om}{4\pi i}
\frac{\VVp(\om)}{\nu - \om}\hat G_{2n+1}(\om) - W_0(\nu)\hat G_{2n+1}(\nu)
\label{hatrecurrence}
\eea
while that of \eq{2smart} is
\be
\sum_k k t_k \hat G_{2k-1}(\l) = - c\sigma [\l W_0(\l) -1 ]~.
\label{hatsmart}
\ee

Notice that Eqs.~\rf{hatmain1} and \rf{hatmain2} (or \rf{hatrecurrence} and
\rf{hatsmart}) express $W(\nu,\la)$ and $\hat G(\nu,\la)$
via $\VVp(\om)$ (or $W_0(\om)$)
which is considered as given. To find it for given $V^\p(\om)$ and,
therefore, to determine $F(\om)$, one can use the equation
\be
\hat G_{1}(\nu)=-\sigma \int_{C_1} \frac{d \om}{2\pi i} \frac{F(\om)}{\nu -
\om}W_{0}(\om)
\label{F}
\ee
which results from the definitions~\rf{defG}, \rf{hatdefG}, \rf{hatbcG} and
\eq{Lambda}.
Using~\rf{hatrecurrence} \eq{F} can be rewritten as
\be
\int_{C_1} \frac{d \om}{4\pi i}
\frac{\VVp(\om)}{\nu - \om}W_{0}(\om) +\left(\frac {s+1}{\nu} - W_0(\nu)
\right) W_{0}(\nu)
   ={\sigma}{c} \int_{C_1} \frac{d \om}{2\pi i}
\frac{F(\om)}{\nu - \om}W_{0}(\om)
\label{FW}
\ee
which looks similar to the large-$N$ loop equation~\rf{1le} for the fermionic
one-matrix model. Thus, one concludes that
\be
2 \Co W_0(\nu) = \VVp(\nu) - 2\sigma c F(\nu) + \frac{s+1}{\nu}
\label{onematrix}
\ee
at the cut (or cuts) of $W_0$.  Eqs.~\rf{defL}, \rf{hatsmart} and
\rf{onematrix} unambiguously fix $\VVp(\om)$, $F(\om)$ and $W_0(\om)$
in the full analogy~\cite{DMS93} with the Kazakov--Migdal model.
For Kogut--Susskind fermions when $s=1$ and $\sigma=-1$, these equations
coincide with those of the fermionic two-matrix model discussed in the previous
section.

\subsection{Explicit solution for quadratic potential}

The analysis for the case of the quadratic potential $V(\om)=m\om^2$ is similar
to that of Sect.~\ref{2exact}.  The solution to Eqs.~\rf{defL}, \rf{hatsmart}
and \rf{onematrix} reads
\be
W_{0}(\l)=\frac{1}{2}
\left[\mu\l+\frac{s+1}{\l}-\frac{1}{\l}
\sqrt{\mu^{2}\l^{4}+2(s-1)\mu\l^{2}+(s+1)^{2}}\right]
\label{DGauss}
\ee
 with
\be
\mu=\frac{(D-1)m+D\sqrt{m^{2}-4 \sigma c^{2}(2D-1)}}{(2D-1)}~.
\label{mu}
\ee
Analogously, one gets
\be
F(\om)=\frac{2 c}{\sqrt{\mu^{2}+4\sigma c^{2}}+\mu}\,\om
\label{Fquadra}
\ee
or substituting \eq{mu}
\be
F(\om)=\frac{2 c}{\sqrt{m^{2}-4\sigma c^{2}(2D-1)}+m}\,\om\,.
\label{Fquadr}
\ee

\eq{Fquadr} coincides with the result of Ref.~\cite{KhM92b} where the
same formula was obtained utilizing the analogy between adjoint and fundamental
fermions in the phase with local confinement for the quadratic potential.
Similarly, picking up the $1/\l^3$ term in \eq{DGauss}, one gets
\be
\LA \ntr \bar\Psi \Psi\RA = - \frac s\mu \,,
\label{PP}
\ee
where $\mu$ is given by \eq{mu},
which agrees with the result of Ref.~\cite{KMNP81} for
lattice QCD with fundamental fermions
at vanishing plaquette term.
Since \rf{PP} is nonvanishing at $m=0$, chiral symmetry is spontaneously
broken.

The fact that $\LA \ntr \bar\Psi \Psi\RA$ coincides for adjoint and fundamental
fermions is based on the sum-over-path representation which is discussed in
Appendix~B. This coincidence is no longer valid for higher correlators, say
$\LA \ntr (\bar\Psi \Psi)^k\RA$ with $k\geq2$ does not coincide with the proper
correlator for the fundamental representation. The value of this correlator
for AFM can be obtained picking up the $1/\l^{2k+1}$ of the expansion
of~\rf{DGauss}.

\newsection{Conclusions}

The main result of this paper is that the fermionic matrix models can be solved
by the method of loop equations quite similarly to the Hermitean ones despite
the fact that Grassmann matrices can not be diagonalized. We have obtained the
explicit solution for case of the simplest quadratic potential in the phase
with local confinement.
It is doubtful, however,  that a complete set of equations can be
 written down for only $G(\nu,\l)$ and $W(\nu,\l)$, which are defined by
 Eqs.~\rf{defG} and \rf{defH}, in the phase with area law.

The fact that the critical index $\gamma_{string}$ of the fermion one-matrix
model coincides in genus zero with the one of the Hermitean one-matrix model
does not necessarily mean that the two models are identical in the continuum.
An example is the model of Ref.~\cite{open} which is associated with $D=0$
open strings. In contrast to the standard case, one might expect
alternating signs of the genus expansion of the partition function
of the fermionic one-matrix model which makes it convergent, since the
Grassmann integral in~\rf{1partition} is always convergent.
It is hardly to imagine, however, that the double scaling limit of the
fermionic one-matrix model differs from that of Ref.~\cite{ds} by something
except signs. This continuum limit of the fermionic one-matrix model deserves
future investigations.

\eop

\setcounter{section}{0}
\setcounter{subsection}{0}

\appendix{Fermion versus Hermitean one-matrix models
\label{log}}

We demonstrate in this appendix the equivalence of the fermion one-matrix model
and the Hermitean one with a logarithm added to the potential.
We start from the derivation of the loop equations for the
Hermitean one-matrix model with the logarithmic potential
\be
Z^{\bf H}=\int d\Phi \e^{-N\tr{} U(\Phi)}
\label{logpartition}
\ee
where
\be
U(\Phi)=V^{\bf H}(\Phi)+\a\log\Phi~,~~~~~
{V}^{\bf H}(\Phi)=\sum_{k=0}^{\infty}g_{k}\Phi^{k}
\ee
and $\Phi$ is $N\times N$ Hermitean.
The models of this type were used to calculate the virtual Euler
characteristics of the moduli space of Riemann surfaces \cite{Pen87} and
are known to describe $c=1$ matter compactificated on a circle of the
self-dual radius interacting with $2D$ gravity~\cite{DV91}. The large-$N$
limit of these models was studied in Ref.~\cite{Tan91}.

To avoid the divergence at $\Phi=0$, let us consider the loop equation
which results from an arbitrary shift of $\Phi$ in
\be
\LA \ntr{} \Big( t^A \frac{\Phi}{\l - \Phi} \Big) \RA = 0
\ee
where the average is with the same measure as in~\rf{logpartition}.
Introducing the loop correlator
\be
{W}^{\bf H}(\l)\equiv\left\langle \ntr{\frac{1}
{\l-\Phi}} \right\rangle \,,
\label{logloop}
\ee
one writes the loop equation as follows
\be
\int_{{C}_{1}}\frac{d\om}{2\pi
i}\frac{\om {V'}^{\bf H}(\om)}{\l-\om}{W}^{\bf H}(\om)=\l
\Big({W}^{\bf H}(\l)\Big)^{2}-
\a {W}^{\bf H}(\l)+\frac{1}{N^{2}}\frac{\delta}{\delta {V}^{\bf H}(\l)}
{W}^{\bf H}(\l),
\label{logloopeq}
\ee
where the contour ${C}_{1}$ encircles all singularities of
${W}^{\bf H}(\l)$ and
\be
\frac{\delta}{\delta {V}^{\bf H}(\l)}=-\sum_{k=0}^{\infty}\l^{-k-1}
\frac{\partial}
{\partial g_{k}}
\label{loginsertion}
\ee
is the loop insertion operator.

\eq{logloopeq} coincides for $\a=2$
with the loop equation~\rf{1loopeq} of the fermion
one-matrix model providing
\be
 \l W^{\bf H}(\l^2) = W(\l)
\ee
where  $W^{\bf H}(\l^2)$ and $W(\l)$ are defined by \eq{logloop} and
\eq{multiloop}, respectively, while
\be
{V}^{\bf H}(\om^2) = V(\om)\,.
\label{VVVV}
\ee
This proves the equivalence of the fermion one-matrix model and the Hermitean
one with the logarithmic potential to any order in $1/N$.

This equivalence can be formally shown at the level of the partition
functions~\rf{1partition} and \rf{logpartition}. To this aim, let us
rewrite~\rf{1partition} as%
\footnote{Notice that if $\Psi$ and $\bar{\Psi}$ were just {\it complex}\/
rather than Grassmann matrices, the determinants would cancel while $\Phi$
would be positive definite.}
\be
Z=\int d\Psi d\bar{\Psi} d \Phi \e^{-N\tr{}V(\sqrt{\bar{\Psi}\Psi})} \,
\delta(\Phi - \bar{\Psi}\Psi) \propto
\int d \Phi \e^{-N\tr{} V(\sqrt{\Phi})} \Big( \hbox{det} \Phi \Big) ^{-2N}
\label{eqpartition}
\ee
which coincides with~\rf{logpartition} providing $\a=2$ and \eq{VVVV} holds.
The matrix $\Phi$ is to be identified with $\bar{\Psi}\Psi$.
The integral~(\ref{logpartition}) is, however, ill-defined  for finite $N$
 because of the divergence at zero.

The equivalence discussed in this appendix makes it possible to understand why
the explicit $N=\infty$ formulas of Subsect.~\ref{1sol} look quite similar to
the proper formulas for the Hermitean one-matrix model (see, {\it e.g.},
Ref.~\cite{Mak91}). In particular, the chiral-noninvariant case when $V$
involves only odd powers of $(\bar{\Psi}\Psi)$ is associated with the Hermitean
one-matrix model with an odd potential ${V}^{\bf H}(\l)=-{V}^{\bf H}(-\l)$.

\appendix{Contraction of the one-link correlator}

Let us consider the following correlator
\be
G^i_{\mu}({\cal O}) \equiv
\left\langle\ntr{}\left({\cal O}\,U_{\mu}(x)\Psi^{i}(x+\mu)U_{\mu}
^{\dagger}(x)\right)\right\rangle,
\label{corr}
\ee
where $\cal{O}$ does not depend on $U_{\mu}(x)$, $U_{\mu}^{\dagger}(x)$,
 $\bar{\Psi}(x+\mu)$ and $\Psi(x+\mu)$, but is arbitrary in other respects.
The averaging is determined by the partition function
(\ref{fpartition}). We shall show that
the calculation of this correlator within the large mass expansion can be
reduced in the large-$N$ limit
to the calculation of the correlator which is independent of $U_{\mu}(x)$,
$U_{\mu}^{\dagger}(x)$, $\bar{\Psi}(x+\mu)$ and $\Psi(x+\mu)$:
\be
\left\langle\ntr{}\left({\cal O}\,U_{\mu}(x)\Psi^{i}(x+\mu)U_{\mu}
^{\dagger}(x)\right)\right\rangle=(P_{\mu}^{+})^{ij}\left\langle
\ntr{}\left({\cal O}\,\frac
{F(\sqrt{\bar{\Psi}(x)\Psi(x)})}{\sqrt{\bar{\Psi}(x)\Psi(x)}}
\,\Psi^{j}(x)\right)\right\rangle,
\label{contraction}
\ee
where $F(\om)$ is some universal (\ie $\cal{O}$-independent) odd
 analytic function
 \be
 F(\om)=\sum_{n=1}^\infty F_n \om^{2n-1}.
 \label{Fdef}
 \ee

 If we expand the exponent in (\ref{fpartition}) in the power series in
 all couplings except the mass of the fermion field then the calculation
 of (\ref{corr}) reduces to the Wick contractions among $\bar{\Psi}(y)$
 and $\Psi(y)$ and group integration over all $U_{\mu}(y)$. The crucial point
 is that the group
 integration goes independently on each link
 due to the absence of the kinetic term of the gauge field. So it is
 sufficient to prove (\ref{contraction}) for the one-link
  correlator~\rf{1link}.  The remaining part of the large mass expansion
  decouples because of the properties of $U(N)$ group integrals and the
  large-$N$ factorization.  Although the coefficients $F_{n}$ will be
  different
 for (\ref{1link}) and for the full lattice correlator (\ref{corr}).

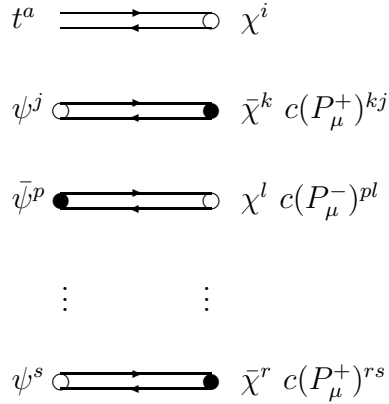
\begin{figure}[tbp]
\unitlength=1.00mm
\linethickness{0.5pt}
\begin{picture}(33.00,77.00)(-70,40)
\put(20.00,105.00){\vector(-1,0){11.00}}
\put(9.00,105.00){\line(-1,0){9.00}}
\put(-6.5,105){$t^{a}$}
\put(0.00,107.00){\vector(1,0){11.00}}
\put(11.00,107.00){\line(1,0){9.00}}
\put(20.00,106.00){\circle{2.00}}
\put(24,105){$\chi^{i}$}
\put(0.00,95.00){\vector(1,0){11.00}}
\put(11.00,95.00){\line(1,0){9.00}}
\put(0.00,94.00){\circle{2.00}}
\put(20.00,94.00){\circle*{2.00}}
\put(-6.5,93){$\psi^{j}$}
\put(24,93){$\bar{\chi}^{k}\,\,c(P_{\mu}^{+})^{kj}$}
\put(20.00,93.00){\vector(-1,0){11.00}}
\put(9.00,93.00){\line(-1,0){9.00}}
\put(0.00,83.00){\vector(1,0){11.00}}
\put(11.00,83.00){\line(1,0){9.00}}
\put(0.00,82.00){\circle*{2.00}}
\put(20.00,82.00){\circle{2.00}}
\put(-6.5,81){$\bar{\psi}^{p}$}
\put(24,81){$\chi^{l}\,\,c(P_{\mu}^{-})^{pl}$}
\put(20.00,81.00){\vector(-1,0){11.00}}
\put(9.00,81.00){\line(-1,0){9.00}}
\put(0.00,59.00){\vector(1,0){11.00}}
\put(11.00,59.00){\line(1,0){9.00}}
\put(0.00,58.00){\circle{2.00}}
\put(20.00,58.00){\circle*{2.00}}
\put(-6.5,57){$\psi^{s}$}
\put(24,57){$\bar{\chi}^{r}\,\,c(P_{\mu}^{+})^{rs}$}
\put(20.00,57.00){\vector(-1,0){11.00}}
\put(9.00,57.00){\line(-1,0){9.00}}
\put(0,70){\makebox(0,0)[lc]{$\vdots$}}
\put(20,70){\makebox(0,0)[rc]{$\vdots$}}
\end{picture}
\caption[x]   {\hspace{0.2cm}\parbox[t]{13cm}
{\small
   The graphic representation of the large mass expansion of the one-link
   correlator~\rf{1link}. }}
\label{fig.1}
\end{figure}

 It is easy to show that (\ref{contraction}) holds for (\ref{1link}). The
 graphic representation of the large mass expansion of (\ref{1link}) for
 the Gaussian potential is depicted in Fig.~\ref{fig.1}.
 The integration over
 $\bar{\chi}$, $\chi$ can be done using the Wick rules (see Fig.~\ref{fig.2}),
\begin{figure}[tbp]
\unitlength=1.00mm
\linethickness{0.5pt}
\begin{picture}(70.00,65.00)(-40,52)
\put(0.00,82.00){\circle*{2.00}}
\put(0.00,70.00){\circle{2.00}}
\put(0.00,94.00){\circle{2.00}}
\put(20.00,105.00){\vector(-1,0){11.00}}
\put(9.00,105.00){\line(-1,0){9.00}}
\put(-6.5,106){$t^{a}$}
\put(0.00,107.00){\vector(1,0){11.00}}
\put(11.00,107.00){\line(1,0){9.00}}
\put(0.00,95.00){\vector(1,0){11.00}}
\put(11.00,95.00){\line(1,0){9.00}}
\put(20.00,93.00){\vector(-1,0){11.00}}
\put(9.00,93.00){\line(-1,0){9.00}}
\put(20.00,100.00){\oval(10,10)[r]}
\put(20.00,100.00){\oval(14,14)[r]}
\put(0.00,83.00){\vector(1,0){11.00}}
\put(11.00,83.00){\line(1,0){9.00}}
\put(20.00,81.00){\vector(-1,0){11.00}}
\put(9.00,81.00){\line(-1,0){9.00}}
\put(0.00,71.00){\vector(1,0){11.00}}
\put(11.00,71.00){\line(1,0){9.00}}
\put(20.00,69.00){\vector(-1,0){11.00}}
\put(9.00,69.00){\line(-1,0){9.00}}
\put(20.00,76.00){\oval(10,10)[r]}
\put(20.00,76.00){\oval(14,14)[r]}
\put(29,87){$=2\,\frac{\sigma
c^{3}}{m^{2}}\,\,\frac{1}{N}\tr{}\bar{\psi}\psi
\,\,(P_{\mu}^{+})^{ij}\,\frac{1}{N}\tr{}t^{a}\psi^{j}$}
\end{picture}
\caption[x]   {\hspace{0.2cm}\parbox[t]{13cm}
{\small
   An example of the use of the Wick rules for integration over $\bar{\chi}$,
   $\chi$ in the Gaussian model.}}
   \label{fig.2}
   \end{figure}
   after which the group
 integration is trivial, because all $U$ matrices cancel with $U^{\dagger}$.
 The answer has the form (\ref{contraction}) with $F(\om)=F_{1}\om$ due to the
 properties of projectors. The value of $F_{1}$ for the $D$-dimensional model
 is given by (\ref{Fquadr}). For the non-Gaussian potential the diagrams of the
 type depicted on Fig.~\ref{fig.3}
 \begin{figure}[tbp] 
 \unitlength=1.00mm
   \linethickness{0.5pt}
   \begin{picture}(70.00,65.00)(-40,52)
\put(0.00,82.00){\circle*{2.00}}
\put(0.00,70.00){\circle{2.00}}
\put(0.00,94.00){\circle{2.00}}
\put(20.00,105.00){\vector(-1,0){11.00}}
\put(9.00,105.00){\line(-1,0){9.00}}
\put(-6.5,106){$t^{a}$}
\put(0.00,107.00){\vector(1,0){11.00}}
\put(11.00,107.00){\line(1,0){9.00}}
\put(0.00,95.00){\vector(1,0){11.00}}
\put(11.00,95.00){\line(1,0){9.00}}
\put(20.00,93.00){\vector(-1,0){11.00}}
\put(9.00,93.00){\line(-1,0){9.00}}
\put(20.00,100.00){\oval(10,10)[r]}
\put(0.00,83.00){\vector(1,0){11.00}}
\put(11.00,83.00){\line(1,0){9.00}}
\put(20.00,81.00){\vector(-1,0){11.00}}
\put(9.00,81.00){\line(-1,0){9.00}}
\put(0.00,71.00){\vector(1,0){11.00}}
\put(11.00,71.00){\line(1,0){9.00}}
\put(20.00,69.00){\vector(-1,0){11.00}}
\put(9.00,69.00){\line(-1,0){9.00}}
\put(20.00,76.00){\oval(10,10)[r]}
\put(20.00,88.00){\oval(10,10)[r]}
\put(20.00,88.00){\oval(14,38)[r]}
\put(29,87){$=-4\,\frac{\sigma
c^{3}g}{m^{3}}\,\,(P_{\mu}^{+})^{ij}\,\frac{1}{N}\tr{}t^{a}\psi^{j}
\bar{\psi}\psi$}
\end{picture}
\caption[x]   {\hspace{0.2cm}\parbox[t]{13cm}
{\small
   An example of the contraction of color indices in the theory with
   quartic interaction: $g$ is the coupling constant. }}
   \label{fig.3}
     \end{figure}
 appear, but their contributions are also of the form (\ref{contraction}),
 (\ref{Fdef}), \ie the terms like $\ntr{}(t^{a}\bar{\psi}\bar{\psi}
 \psi\psi)$ do not arise. This is because the potential depends only on
 $\bar{\psi}\psi$.

 In the case of Kazakov-Migdal model the formula (\ref{contraction}) holds
 even if the averaging in (\ref{1link}) is only w.r.t.\ $U_{\mu}(x)$
 \cite{Mig92a,Mig92d}. This result is nonperturbative (it follows simply
 from the gauge invariance and the large-$N$ factorization). Also it is
 possible to calculate the pair correlator of the gauge field provided
 that all the correlators of the type $\langle \ntr{}\Phi^{n}\rangle$
 are known \cite{DMS93}. This is {\it not}\/ the case for the model under
 consideration. The reason is that one needs to compute a more complete set
 of correlators than $\langle \ntr{}(\bar{\Psi}\Psi)^{n}\rangle$, for
 example the correlator of the type $\langle\ntr{}(\bar{\Psi}\bar{\Psi}
 \Psi\Psi)\rangle$.

 Of course, our proof of \eq{Lambda}
 is valid only in the local confinement phase which is
 characterized by vanishing expectation values of nontrivial adjoint Wilson
 loops
 \be
 W_{A}(C)\equiv\left\langle\left|\ntr{}U(C)\right|^{2}\right\rangle
 =\delta_{0,A_{min}(C)},
 \label{locconf}
 \ee
 where $A_{min}(C)$ is area of the minimal  surface spanned by $C$.
 Eq. (\ref{locconf}) is valid to all orders of the large mass expansion
 and was implicitly used in the proof.

\begin{figure}[tbp]
\begin{picture}(120.00,220.00)(-30,125)
\unitlength=1.00mm
\linethickness{0.5pt}
\put(35.00,61.00){\oval(18.00,2.00)[r]}
\put(30.00,62.00){\line(0,1){14.00}}
\put(28.00,60.00){\line(0,1){18.00}}
\put(28.00,60.00){\line(1,0){7.00}}
\put(30.00,76.00){\line(1,0){14.00}}
\put(28.00,78.00){\line(1,0){14.00}}
\put(44.00,76.00){\line(0,1){20.00}}
\put(42.00,78.00){\line(0,1){18.00}}
\put(82.00,61.00){\oval(22.00,2.00)[l]}
\put(83.00,62.00){\oval(2.00,16.00)[t]}
\put(83.00,60.00){\oval(2.00,28.00)[b]}
\put(84.00,62.00){\line(1,0){14.00}}
\put(84.00,60.00){\line(1,0){16.00}}
\put(98.00,62.00){\line(0,1){50.00}}
\put(100.00,89.00){\oval(28.00,2.00)[r]}
\put(100.00,60.00){\line(0,1){28.00}}
\put(100.00,90.00){\line(0,1){22.00}}
\put(42.00,97.00){\oval(78.00,2.00)[l]}
\put(84.00,96.00){\line(0,1){16.00}}
\put(84.00,112.00){\line(1,0){14.00}}
\put(44.00,96.00){\line(1,0){40.00}}
\put(82.00,98.00){\line(-1,0){40.00}}
\put(100.00,113.00){\oval(68.00,2.00)[r]}
\put(82.00,98.00){\line(0,1){16.00}}
\put(82.00,114.00){\line(1,0){18.00}}
\put(85.00,97.00){\makebox(0,0)[lc]{$x+\mu$}}
\put(67.00,101.00){\makebox(0,0)[cb]{$x$}}
\put(68.00,98.00){\circle*{1.0}}
\put(82.00,98.00){\circle*{1.0}}
\put(76.50,103.00){\makebox(0,0)[cb]{$L_{x,x+\mu}$}}
\put(76.50,91.00){\makebox(0,0)[ct]{$\Gamma_{x+\mu,x}$}}
\put(30.00,62.00){\line(1,0){5.00}}
\end{picture}
\caption[x] {\hspace{0.2cm}\parbox[t]{13cm}
{\small
   The typical paths $\Gamma_{x+\mu,x}$ which contribute to the sum
   on the r.h.s.\ of \eq{sumoverpath} in the local confinement phase. These
   $\Gamma_{x+\mu,x}$ coincide with $L_{x,x+\mu}$ passed backward modulo
   backtrackings which form a $1D$ tree.  The sum over $\Gamma_{x+\mu,x}$ is
   reduced to summing over the backtrackings.}}
   \label{fig.4}
   \end{figure}
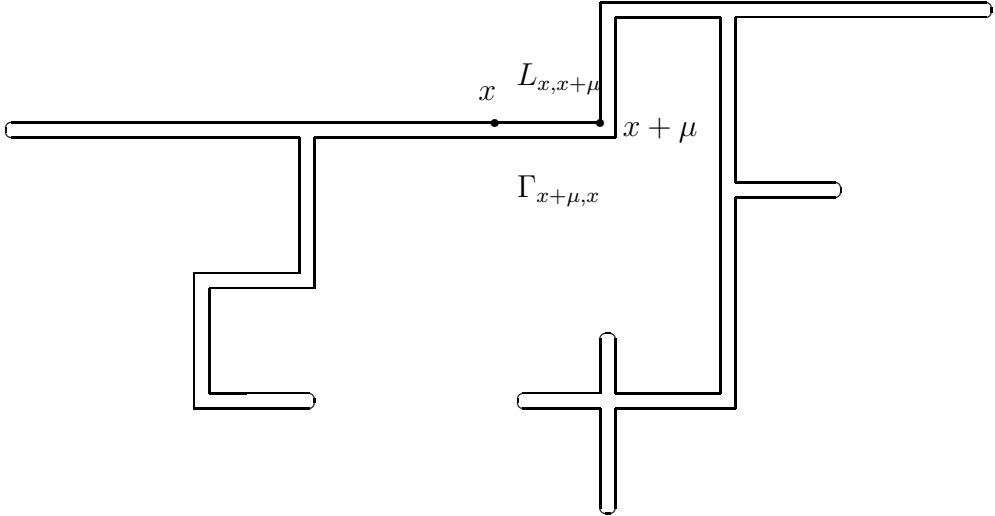

 To demonstrate it, let us consider the following
 example. The well-known sum-over-path representation of the correlator
 \be
 \left\langle\ntr{}\bar{\Psi}(x)U_{\mu}(x)\Psi(x+\mu)U_{\mu}^
 {\dagger}(x)\right\rangle=-\sum_{\Gamma_{x+\mu,x}}\frac{W_{A}(L_
 {x,x+\mu}\Gamma_{x+\mu,x})}{m^{|\Gamma_{x+\mu,x}|+1}}\Sp
 P^{-}(\Gamma_{x+\mu,x}),
 \label{sumoverpath}
 \ee
 where $L_{x,x+\mu}$ is the link connecting $x$ with $x+\mu$ and the sum
 goes over all paths $\Gamma_{x+\mu,x}$ complementing $L_{x,x+\mu}$
 for a closed contour, $P^{-}(\Gamma_{x+\mu,x})$ being the product of
 $P_{\mu}^{-}$ ordered along this path.
 In the phase with local confinement one substitutes~\rf{locconf}
 in this formula so that the typical paths $\Gamma_{x+\mu,x}$
 coincide~\cite{KhM92b} with $L_{x+\mu,x}$ modulo backtrackings as is depicted
 in Fig.~\ref{fig.4}.
 In the phase with normal area law, nontrivial loops contribute to the sum
 in~\rf{sumoverpath}
 which were disregarded in our
 consideration of the group integrals. It is doubtful that the formula
 (\ref{contraction}), which enables one to write down a complete set of
 Schwinger--Dyson equations for local objects, really works in the area law
 phase where the dynamics
 of some extended objects is expected to be nontrivial.



\end{document}